# Critical phenomena in plasma membrane organization and function

Thomas Shaw[1], Subhadip Ghosh[2], Sarah L. Veatch[1,2,*]

**Abstract:** Lateral organization in the plane of the plasma membrane is an important driver of biological processes. The past dozen years have seen increasing experimental support for the notion that lipid organization plays an important role in modulating this heterogeneity. Various biophysical mechanisms rooted in the concept of liquid-liquid phase separation have been proposed to explain diverse experimental observations of heterogeneity in model and cell membranes, with distinct but overlapping applicability. In this review, we focus on evidence for and consequences of the hypothesis that the plasma membrane is poised near an equilibrium miscibility critical point. Critical phenomena explain certain features of the heterogeneity observed in cells and model systems, but also go beyond heterogeneity to predict other interesting phenomena, including responses to composition perturbations.

## 0. Overview

Spatial organization of the plasma membrane on 10-100 nm length scales has been a topic of interest to biology for decades. Heterogeneity has been hypothesized to play important roles in many membrane-associated biological processes, from coordination of signal transduction machinery to endo- and exocytosis to polarization, as iconically described in the lipid raft hypothesis (1). Various biophysical notions have been marshaled to provide explanations of membrane heterogeneity, especially from equilibrium thermodynamics and phase transitions, as coexisting liquid phases are readily observed in both purified membranes and membranes isolated from plasma membranes (2, 3). One important thread of this ongoing conversation explains nanoscale membrane structure as critical phenomena. In contrast to a classical phase separation picture of plasma membrane domains, which implies stable and well-defined discrete regions of defined composition, critical phenomena are subtle, dynamic and malleable, and inhabit the relevant nanoscopic length-scales.

[1]Program in Applied Physics, University of Michigan, [2]Program in Biophysics, University of Michigan, *Correspoding author, contact: sveatch@umich.edu

In this review, we conduct a brief historical survey of membrane domains in both model and biological membranes and describe the consensus that has been reached regarding the macroscopic miscibility phase behavior of model membranes as well as remaining controversies regarding the microscopic heterogeneity reported in other regions of phase space. We introduce membrane criticality and the accumulated evidence that eukaryotic plasma membranes are near-critical. We discuss where the concept of criticality fits into conventional descriptions of raft phenomenology, and describe some unique areas where criticality could plays roles in biological function that go beyond simply organizing components.

**1. An early history of domains in model and cell membranes.**

There is a long history of detecting heterogeneity in bilayer membranes containing cholesterol. Some of the earliest studies that interrogated purified membranes in the late 1960s and early 1970s found evidence that liquid membranes containing a single phospholipid species and cholesterol contained structure on 1-100nm length-scales accessible to the spectroscopic methods available at the time (4–9). Over the decades, evidence for these microscopic domains in binary mixtures of phospholipids and cholesterol continued to accumulate, through additional spectroscopic studies, calorimetry, the application of fluorescence techniques such as fluorescence quenching, anisotropy, and Förster resonance energy transfer (FRET), and imaging methodologies such as freeze fracture electron microscopy (10–21).

Largely in parallel to this physical chemistry characterization of membranes, cell biologists also postulated that membranes within cells could contain lipid-mediated sub-structures that could be important for cell functions. Early evidence came from the observation that cells could polarize their membrane lipid composition via differential trafficking of lipid species (22). It has also long been appreciated that cholesterol or similar sterols are vital to the proper functioning of most eukaryotic plasma membranes, as manipulation of membrane cholesterol content was shown to interfere with functions including endocytosis, several receptor mediated signaling cascades, and cell cycle control (23–26). In the early 1990s it was discovered that some detergents only partially solubilize cellular membranes, especially the plasma membrane (27, 28), and it was postulated that the insoluble fractions represented distinct membrane domains (1, 20). Researchers found that artificially clustering some membrane components individually could drive macroscopic colocalization, termed co-patching, detectable by conventional fluorescence microscopy (29). Subsequent spectroscopic and fluorescence measurements supported the hypothesis that cell membranes were heterogeneous in their lipid and protein content (30–32), and many studies correlated biological function with presumed membrane heterogeneity probed and perturbed via these and related assays (33–36). Each of these methods had well documented flaws (37), but the remarkable consistency of the conclusions drawn from different methodologies provided convincing evidence that lipid-driven heterogeneity is relevant to the plasma membrane.

**2. Liquid-liquid phase separation in model membranes.**

***Purified membranes:***

In 2001, the first observations of macroscopically phase separated fluid domains were reported in bilayer membranes reconstituted from cellular extracts (38, 39). These initial observations emphasized similarities between the composition of the coexisting phases in model membranes and heterogeneity measured in cells by detergent solubilization methods. Soon after, it was appreciated that at least three lipid components were required to form macroscopically phase separated domains in bilayers: a high melting temperature lipid, a low melting temperature lipid, and a sterol such as cholesterol (40). Phase diagrams describing this macroscopic phase transition have now been mapped by numerous groups using a range of methods and are in good qualitative agreement (41–49). Since three components are required to observe liquid-immiscibility, phase diagrams have been mapped by varying temperature and the molar fraction of all three components. An introduction to reading and interpreting these phase diagrams is given in Figure 1, and several experimental phase diagrams are presented in Figure 2.

A range of experimental (50–56) and simulation (57, 58) approaches using different lipid combinations have produced results consistent with phase diagrams topologically similar to those shown in Figure 1b, and their detailed characteristics have been described in several comprehensive review articles (3, 59, 60). Liquid immiscibility is most often observed at temperature below the melting temperature ($T_m$) of the high $T_m$ lipid (42, 44–46), although there are exceptions (41). In the typical case, reported phase diagrams have a region of liquid-liquid coexistence, a region of three-phase coexistence (2 liquids and a solid), and two regions of liquid-solid coexistence. One of the liquid phases is called the liquid-disordered (Ld) phase, and it resembles the liquid crystalline phase of pure phospholipids (61). The second liquid phase is called liquid-ordered (Lo) (15), which was first characterized in mixtures of saturated phospholipids and cholesterol (8, 16, 18). The solid phase is often called gel, and is most likely Lβ' (62).

The high cholesterol edge of the three-phase triangle is sloped such that the Lo phase contains a higher cholesterol mole fraction than the Ld phase. This edge of this triangle is also the first tie-line in the liquid-liquid-coexistence region. As cholesterol is increased further, tie-lines run roughly parallel to one another, meaning that cholesterol concentration increases roughly linearly in both phases. The Lo-Ld coexistence region terminates in a miscibility critical point, where in principle tie-lines merge into a single point. In practice this region of the phase diagram is surprisingly flat, meaning that tie-lines remain long and shorten over a very small range of compositions. As temperature is lowered, the Lo-Ld immiscibility gap extends to higher concentrations of cholesterol and low Tm lipid, as does the concentration of components at the critical point (41, 45, 48). At constant temperature, the miscibility gap expands when the Tm of the high Tm component is increased or when the Tm of the low Tm lipid is decreased (41, 42, 44). No macroscopic miscibility gap is observed for some combinations of low and high Tm lipids (42, 44, 63). A closed loop miscibility gap is found when the extremely low Tm lipid DiPhytanoyl PC is used (41), meaning that Lo-Ld coexistence occurs at temperatures above the Tm of the saturated component and there are two critical points. The phase behavior of the mixed system can depend on more than just the Tm of components. For example, sphingomyelin (SM) lipids are more effective at establishing coexisting phases as compared to glycerol-phospholipid lipids with PC headgroups (44, 58) even when the main chain transition occurs at similar temperatures for SM and PC lipids used.

Although the equilibrium thermodynamics description of the macroscopic miscibility transition is now largely accepted, questions remain regarding the thermodynamic basis of the microscopic heterogeneity also routinely observed in membranes containing cholesterol. These structures are frequently reported at temperatures and compositions where membranes remain uniform on a macroscopic scale using methods sensitive to molecular-scale organization such as FRET, fluorescence quenching, or ESR. Membranes can be tuned from a state with macroscopic phase separation to one with microscopic heterogeneity by raising temperature (64, 65), titrating in an additional component that disrupts the macroscopic phase transition (e.g. (50, 52, 66)), or by probing different ratios of the same lipid species at fixed temperature (65). Submicron structure is also reported in binary mixtures of saturated lipids and cholesterol (7, 67) and in some ternary membranes that do not exhibit macroscopic Lo-Ld immiscibility at any temperature or lipid ratio (21, 47, 63, 68). Recent experimental developments in purified membranes has begun to probe how leaflet asymmetry impacts phase separation and the presence of submicron structure in purified membranes (69, 70).

Experimental observations have motivated numerous theories to explain the presence of microstructure at thermodynamic equilibrium. Critical phenomena provide a possible mechanism to bridge macro- and micro-scales in the form of dynamic fluctuations (45, 71). Other theories enable static, finite-sized domains by including some repulsive mechanism to oppose formation of macroscopic domains (72–74). These theories too predict domains that span macro-to micro- scales. Recent experimental work has begun to directly test some of these theories explicitly (75).

*Isolated membranes:* The first observations of macroscopic liquid immiscibility in vesicles blebbed directly from living cells came in 2007 (2), and these vesicles were named Giant Plasma Membrane Vesicles (GPMVs) due to their close resemblance to the Giant Unilamellar Vesicles (GUVs) used widely in fluorescence microscopy investigations of purified lipid mixtures. Earlier work using a similar vesicle preparation had characterized their lipid and protein content by mass spectrometry (76), and had observed heterogeneity using ESR (30), a spectroscopic method that can detect heterogeneity on the molecular scale. Baumgart et al (2) detected the sorting of fluorescent lipid analogs and fluorescently tagged proteins with respect to phases in GPMVs at temperatures well below those where cells were grown, leading the authors to conclude that this phase transition was not relevant for cells under normal growth conditions.

Soon after, other methods emerged to isolate plasma membranes from cells and all could yield coexisting liquid phases (77, 78), although the conditions needed to achieve phase separation differed between methods used. A subsequent study found correlations between biochemically defined detergent resistant membranes and the liquid-ordered phase detected in GPMVs (79), and that the surface fraction of ordered phase at low temperature was altered by acute treatments to manipulate cholesterol levels in vesicles (79, 80). In all cases, miscibility transition temperatures ($T_{mix}$) remained well below growth temperatures in isolated cells, emphasizing that such macroscopic domains were not likely to form under physiological conditions. A possible explanation came in 2008 when it was shown that freshly isolated GPMVs exhibited hallmarks of criticality, placing them close to a room temperature miscibility critical point (81). Over time, additional studies have documented how GPMV phase behavior is impacted by growth conditions in cells (82–84) and differentiation into other cell types (85, 86), and lipidomics analysis has begun to characterize the vast compositional complexity of these membranes (82, 84, 87).

While there are many similarities between the phase behavior observed in GPMVs and purified model membranes, there are key differences (88). The coexisting phases detected in GPMVs differ in their physical properties from their purified membrane counterparts. The 'fluidity' of phases are more similar in GPMVs compared to GUVs, as measured through diffusion of membrane components or using order sensing fluorophores that report on local hydration within the hydrophobic region of the membrane (89). These different physical properties can be sensed by incorporated proteins. Some transmembrane proteins, particularly those with palmitoylated cysteines, are observed to partition into the Lo phase in GPMVs whereas few are reported to partition into the Lo phase in purified membranes (90).

It should be noted that GPMVs are model membranes that differ from intact plasma membranes in important ways. Plasma membranes exist in close association to the actin cytoskeletal cortex, while GPMVs are missing polymerized cytoskeletal components and tend to be depleted in proteins that associate with the actin (91). Notably, the cell plasma membrane does not macroscopically phase separate even under conditions that cause GPMVs to phase separate, or in any known conditions, even when phase separated GPMVs remain attached to an intact cell membrane (92). GPMVs are depleted of PI(4,5)P$_2$ (93), which typically makes up several mol% of the inner plasma membrane leaflet in intact cells (94). A recent report demonstrates that isolated GPMVs are frequently permeable to large hydrophilic markers (95), indicating that their membranes contain long-lived defects. While cell membranes are asymmetric in their lipid and protein composition, at least some of this asymmetry is lost in the GPMV generation and isolation process (2, 93). The interpretation of GPMV experiments is also complicated by their sensitivity to methodological choices. For example, the most common method to prepare GPMVs involves incubating cells with a low concentration of formaldehyde and a reducing agent, and the choice of the reducing agent can greatly impact the transition temperature and physical properties of phases of the resulting GPMVs (78, 96). This is due, at least in part, to the ability of some reducing agents to modify membrane proteins and lipids.

**3. Phase separated domains are related to but different from 'raft' heterogeneity in intact cells.**

Phases are robust, macroscopic entities with well-defined compositions, and with the exception of the yeast vacuole (97, 98), structures resembling liquid-liquid phase separation are not observed in intact cells. Instead, the vast majority of membrane domains in cells are microscopic and require significant perturbations in order to be visualized or isolated. In line with this, several recent studies have directly concluded that there is no evidence of a miscibility phase transition in intact cells when cells are examined through the lens of several different experimental observables (99, 100). Nonetheless, a large number of experimental studies provide strong evidence that the heterogeneity reported in intact cells is closely related to the macroscopic phase separation observed in GPMVs. Two excellent recent reviews discuss many of these findings, as well as some exceptions (101, 102). We highlight several lines of evidence below.

1. In many cases, proteins that are associated with live-cell heterogeneity or with detergent-resistant membranes are also found to partition into the Lo-like phase of GPMVs. In particular, single-pass transmembrane and peripheral proteins containing palmitoylations are more likely to be found in detergent resistant membranes and partition into the Lo phase in GPMVs, while proteins containing branched and unsaturated geranylgeranyl or prenyl groups tend to be solubilized by detergents and partition into the Ld phase in GPMVs. Recent work begins to extend to multi-pass proteins which can accommodate more complex protein/lipid interactions (103–105). It is likely that protein partitioning will be dictated by the identity of lipids that solvate transmembrane proteins in complex membranes (106, 107)

2. Cells actively tune their plasma membrane $T_{mix}$ in response to changes in growth conditions and conditions that change $T_{mix}$ lead to different phenotypes. For example, experiments indicate that cells in culture actively tune the miscibility phase transition temperature of their membrane to be a fixed temperature below their growth temperature (82), and that $T_{mix}$ lowers in cells under conditions that inhibit cell growth (83). Other studies have documented that acute treatments with lipophilic small molecules or dietary lipids that alter $T_{mix}$ correlate with changes in signaling outcomes, cellular differentiation, and even the general anesthetic response (85, 108, 109). While it is possible that these correlations with $T_{mix}$ are a result of a mutual correlation with an unrelated quantity, the accumulated evidence suggest that maintaining $T_{mix}$ plays important roles in cellular processes.

3. In model membranes, $T_{mix}$ predicts structure observed in single phase membranes. While model membranes appear homogeneous at the macroscopic scale above $T_{mix}$, the value of $T_{mix}$ can predict the presence of structure at smaller scales, or under conditions where membranes are perturbed by clustering one component. A very recent report that documents the temperature dependence of both macroscopic phase behavior and microscopic heterogeneity in GPMVs under a range of perturbation conditions (96). This work found that the microscopic heterogeneity in GPMVs at elevated temperature was highly correlated with their macroscopic transition temperature Tmix, suggesting that the same could be true in intact cells. An older study finds that phase marking probes enrich in membrane domains stabilized through adhesion even well above Tmix (80). These and other studies have led to the idea that the value of Tmix is a measure of 'raft' stability under physiological conditions, even though cells do not appear to experience phase separation directly.

4. Theoretical arguments support that coupling to an intact cytoskeletal network will disrupt the macroscopic phase transition. It is well established that 'quenched disorder' abolishes first order phase transitions in two dimensional systems like membranes (110). The cortical actin cytoskeleton, which is linked to the plasma membrane through a system of adapter proteins, is likely to play the role of quenched disorder in the intact plasma membrane (92, 111). Experimental studies in model membranes support the main conclusions of this theory, that a broadly distributed cytoskeletal network disrupts macroscopic domains while stabilizing small-scale structure (112). Direct tests in intact cells have yet to be reported, but there is broad evidence for important connections between cortical actin and raft heterogeneity (113–117).

5. There is increasing evidence for phase-driven partitioning within intact cells. Recent work utilizing single molecule fluorescence methods are beginning to draw more direct connections between phases in vesicles and domains in cells (118, 119). These studies monitor the recruitment and exclusion of probes with respect to clustered proteins in intact cells and find that probe concentration in clusters mirrors partitioning with respect to phase separated domains in model membranes, although typically with a smaller magnitude.

To explain these experimental developments, several theoretical avenues have been explored, with a goal of explaining the lack of a macroscopic phase separation in intact cells. Any complete biophysical model of membrane heterogeneity must first account for this fact, and therefore must be more complex than the simple phase separation picture. These explanations include coupling to quenched disorder in the form of the cytoskeleton (92, 111), coupling to curvature in a way that produces a microemulsion (72, 74, 120), and nonequilibrium suppression of domain growth, for example by active lipid transport or remodeling of membrane-coupled actin structures (121, 122). Many of these models, including the critical phenomena that are discussed in the next section, are closely related to the canonical phase separation picture (123).

## 4. Criticality and its connection to plasma membrane

Equilibrium critical phenomena are special relationships between thermodynamic properties of a system that emerge due to a diverging correlation length of the system in the vicinity of a critical point. Key physical quantities such as the heat capacity, the susceptibility, and the interfacial energy between domains also exhibit specific behaviors as the critical point is approached. These properties of critical systems are *universal*, meaning that the dominant behavior is governed by a small number of effective parameters, regardless of the complicated details of microscopic interactions (Figure 3). The physics of criticality has been covered in detail in many textbooks (124–127). A useful introduction for biophysicists is given in (128).

The physics of criticality has developed over the last 200 years, beginning with the observation of a critical temperature for several liquids, above which there is no liquid-gas transition (129). The van der Waals equation of state (1873) was the first model that featured a "critical point" with this property (130). Ornstein and Zernicke, in 1918 (131) formalized the study of spatial fluctuations in liquid-gas systems, a key development. In the first half of the 20$^{th}$ century, precise measurements of near-critical liquid-gas systems (132) as well as magnetic systems (133) indicated that classical equations of state were inadequate to explain the near-critical region of phase space, for example that the shape of the phase liquid-gas coexistence curve is qualitatively different than predicted by van der Waals. In 1944, Lars Onsager exactly solved the 2d Ising model, which allowed for the study of critical exponents in that system (134). In 1952, Yang and Lee observed that phase transitions correspond to non-analyticities of the partition function in the thermodynamic limit (135), and Widom and others proposed power-law scaling for various quantities at critical points, and derived relationships between different critical exponents from that hypothesis (136). Real-space (Kadanoff) and momentum-space (Wilson) renormalization group methods explained the emergence of these power laws and provided avenues for computing approximate values for the critical exponents in general systems (137, 138). They also set a uniform framework for other conceptual advances. A broader class of nonequilibrium critical points can also be defined, generalizing this equilibrium concept. Biological nonequilibrium critical points have received considerable attention in recent years, see, e.g. Mora and Bialek for a review (139).

Coexistence regions generically terminate in a critical point except in special circumstances, and the Lo-Ld miscibility of purified and isolated membranes is no exception. On the phase triangle shown in Figure 1B, the critical point occurs on the high-cholesterol edge of the Lo-Ld miscibility gap. Initial evidence of critical behavior in membranes came from NMR studies of multilamellar vesicles, which detected enhanced line-broadening in the vicinity of known critical points (45). This was attributed to the diffusion mediated exchange of lipids between fluctuations with sub-micron dimensions. Later work directly visualized micron-sized critical fluctuations (above the critical temperature; Tc) and fluctuating phases (below Tc) in GUVs of purified lipids, providing evidence that membranes belong to the 2D Ising model universality class, meaning that fluctuations exhibited a temperature dependence that is universal to two dimensional systems with a one-dimensional order parameter (71). Subsequent measurements confirmed this observation in supported membranes by atomic force microscopy (48), and probed the dynamics of critical membranes (140). At this same time, it was discovered that isolated GPMVs also exhibited critical behaviors in the vicinity of a room temperature critical point (81). Again, fluctuations were consistent with the 2D Ising model universality class. These observations and their implications have been reviewed in greater detail previously (128).

One of the key features of a critical point is that its fingerprints extend well beyond the phase transition itself (Figure 3). An important parameter is t, the difference between the temperature of the system and the critical temperature (Tc) normalized by Tc in units of Kelvin. The correlation length $\xi$, or characteristic size of critical compositions fluctuations, is predicted to vary as $\xi(t)=\xi_0/t$, where $\xi_0$ is a parameter with dimensions close to the size of molecules in the system and in membranes was measured to be roughly 1nm. Note that as $T \rightarrow T_c, t \rightarrow 0$, so that the correlation length becomes infinite. Extrapolating this relationship using a room temperature critical point (Tc = 22°C = 295K), then 20nm sized fluctuations are expected at 37°C, which corresponds to t=0.05. This prediction is in good agreement with recent experimental work probing heterogeneity in GPMVs by FRET, which detects evidence for larger than 10nm structure in GPMVs over this same temperature range (96). Similar observations have also been made in purified model membranes (141). Note that *t* need not be a physical temperature. Instead it is any trajectory in the phase diagram that runs perpendicular to tie-lines close to the critical point. Thus, while some Ising model images of Figure 3A are obtained by varying temperature in the model, the corresponding three-component lattice model images are obtained by varying composition at fixed temperature, as indicated in the phase diagram.

Another physical property that can extend well beyond the phase transition itself is the susceptibility ($\chi$). The susceptibility measures how large a local composition difference arises from a local force applied to components of one of the phases, e.g. by clustering components that prefer Lo lipids. In other words, in a highly susceptible membrane, a domain of distinct local composition can be stabilized by only clustering a small subset of components, or by weakly biasing the concentration of many components that have the same order preference. In the Ising universality class, $\chi \propto t^{-7/4}$. This too has experimental support in vesicles, where robust domains are stabilized well above Tmix by organizing a small subset of components by an actin network or streptavidin crystal that partially decorates a vesicle surface (114, 142), or by adhesion to a supported membrane (80).

Direct theoretical predictions of critical phenomena such as the scaling of the correlation length and the magnitude of the susceptibility are quite useful for predicting consequences of perturbations to membrane heterogeneity, but the theory quickly becomes intractable when coupled to more complex biological phenomena. Statistical mechanical lattice models based on the Ising model can be useful in this situation. These models typically only contain two components (up and down 'spins') positioned on a lattice, where the components at the lattice sites are either allowed to change identity (such that the composition or 'magnetization' can vary) or are allowed to exchange with other sites on the lattice (such that the composition remains fixed). Universality guarantees that, so long as the system is close to the critical point, the Ising model captures the relevant mesoscopic heterogeneity of the membrane, for appropriate choices of the Ising reduced temperature t and magnetization m (Figure 3). That is, the Ising model accurately recapitulates the thermodynamics of the effective Lo order parameter at length-scales beyond a few lipid diameters, despite the extreme simplicity of the microscopic interaction in the model – a simple nearest-neighbor interaction potential. As a result, when a biological system is coupled to the Lo order parameter, an Ising model modified to include this coupling is expected to reflect the relevant biophysical phenomena. Past work has used this approach to model the coupling of fluctuations to cortical actin (92), to explain changes in phosphorylation steady states upon clustering of a component, for various values of $t$ and $m$ (118, 143), and to predict how proximity to the critical point affects conformational state equilibria of proteins whose boundaries are sensitive to lipid order (144).

***Evidence for criticality playing a role in cells:*** If intact plasma membranes exhibit similar heterogeneity to that observed in GPMVs, it could easily be relevant to the biological function of membrane proteins. An important line of evidence that criticality plays a role in biological function has come from the tuning of the GPMV critical point. For a system to be near a critical point in the first place, two parameters must be tuned, corresponding to *t* (temperature) and *m* (composition) of the (fixed-composition) Ising model. In the extremely large space of lipid mixtures of varying composition, there are many critical points – an $n-2$-dimensional manifold in the *n*-dimensional space. However, there is no generic reason that tuning the concentration of any given lipid will correspond to tuning just *t*, or just *m*, or neither – general perturbations will affect both *t* and *m*. Thus it is somewhat surprising that the cell arrives near a critical point if it constructs its membranes without explicitly or implicitly tuning to the critical point, given that lipid composition is modulated by a wide variety of perturbations. In other words, it would be surprising if plasma membrane composition is near-critical simply by coincidence. Furthermore, at least in certain cases, eukaryotic cells adapt to perturbations in ways that preserve the distance to the critical point, and corresponding physical properties. Zebrafish cells cultured at a range of temperatures from 20-32 °C produce GPMVs with correspondingly altered Tc (Burns et al 2017).

The concept of a high susceptibility near a critical point is useful in interpreting recent single molecule and super-resolution studies documenting the partitioning of phase marking probes to protein clusters in intact cells (118, 119, 143). In these studies, antibodies are used to cross-link a membrane component that prefers either the Lo or Ld phase, then the differential partitioning of probes is monitored with respect to these domains. When proteins are clustered that themselves prefer the Lo phase, then probes that also prefer Lo tend to be recruited and those that prefer Ld tend to be excluded. In contrast, when proteins that prefer the Ld phase are clustered, then probes that prefer Lo are excluded and probes that prefer Ld are recruited. Similar to experiments with vesicles adhered to supported membranes (80), the act of clustering a protein or peptide biases concentration of many components in ways that can be detected when membranes have high susceptibility. In some cases the extent of probe partitioning approaches that observed in phase separated vesicles (143), while in others the sorting of components is much weaker (118). These differences could arise from differences in the coupling of protein clusters to membranes, or differences in the susceptibility of the membrane in different experimental systems.

## 5. Criticality as it relates to biological function

Since the inception of the raft hypothesis, the functional relevance of membrane domains has focused on their ability to compartmentalize protein and lipid components so that they can optimally function within biochemical networks (1). Critical phenomena are in many ways consistent with this framework. A super-critical membrane contains domains resembling ordered and disordered phases, and components that partition with the same phase will colocalize within these domains. The fluctuations are small and dynamic, consistent with evolving descriptions of rafts over the decades (145–148), but fluctuations alone are not an effective means to strongly colocalize or confine membrane components. This new reality requires us to move beyond the simple mechanisms proposed in the early raft literature to propose and test mechanisms that exploit the unique material properties of critical systems. Several proposals are highlighted in Figure 4 and described below.

***Interactions between proteins:*** Composition fluctuations can mediate forces between membrane proteins, via a process termed 'critical Casimir forces' first described between conducting plates in vacuum (149). In essence, proteins will feel an effective attractive potential if they partition into the same phase because their coming into close proximity allows them to share the same local lipids, as shown in Figure 4A (150, 151). In contrast, proteins that prefer different environments will feel an effective repulsion, since there is an energetic cost to mixing their local environments. These potentials are weak (on the order of the thermal energy $k_BT$) but have a range given by the correlation length and the size of the protein or protein cluster. This is long-ranged compared to other interaction modes experienced by membrane proteins such as curvature, electrostatics, and van der Waals potentials. It is notable that repulsion of components that prefer different phases has a larger magnitude than attraction between components that prefer the same phase. The Casimir force may contribute to the stability of protein assemblies, including phase separated polymer droplets that assemble on membranes. The Casimir force is also expected to alter biochemistry occurring at the membrane, by increasing or reducing the rates at which the proteins encounter one another. It is tempting to speculate that one functional role of palmitoylation, the post-translational modification that places a saturated acyl chain on proteins, is to tune the magnitude of this Casimir force for specific protein species.

***Susceptibility to receptor clustering:*** The high susceptibility of a critical membrane provides a means for the cell to sense a redistribution of a subset of membrane constituents by an external force (Figure 4B). Clustering a membrane protein that prefers one phase would bias the local lipid composition in proportion to the heightened susceptibility of the system. This effect can impact biochemical reactions that take place within these clusters, drastically altering the chemical steady state of the system. We have studied this effect in the context of B cell receptor (BCR) signaling (118, 143). Here, the act of clustering the BCR or another ordered membrane component by an extracellular ligand stabilizes an ordered domain that contains a higher local concentration of kinase and a lower concentration of phosphatase than the membrane as a whole. This establishes a local environment that favors receptor phosphorylation and activation. In principle, this class of activation mechanism could contribute to a wide range of signaling pathways that are initiated by receptor clustering at the cell surface. This type of mechanism could also play a role in establishing biochemical environments in membrane regions where components are organized by processes occurring at the inner plasma membrane leaflet, such as at junctions between the ER and plasma membrane (152), or at sites where scaffolding adaptor proteins are anchored to membranes such as in neuronal synapses (153).

***Allosteric regulation of single proteins:*** Beyond contributing to the organization of proteins, the functioning of single proteins can also be impacted by the size and stability of fluctuations in the membrane. One mechanism that has been proposed requires that two conformational states of the protein in question have different boundary lipid preferences for Lo or Ld lipids (144). If that is the case, then a change in $T_c$ will differentially affect the free energies of the two conformational states. Roughly, a spatially extended lipid preference carries a free energy cost that decreases near $T_c$, so that a conformational state with strong order preference becomes more probable when fluctuations are large compared to the protein diameter. This model was proposed to explain striking correlations between the $T_c$-altering effects and anesthetic or anesthetic-reversing potencies of a wide range of treatments, including short and long-chain n-alcohols and hydrostatic pressure (108, 109).

***Tuning binding of allosteric regulators:*** The chemical potential $\mu$ of a component is the thermodynamic parameter that controls the proclivity of that component to enter or exit a system. For example, its availability to bind in a binding pocket. Formally, the chemical potential is the increment of free energy to move one particle into the system from a particle bath. Equivalently, the chemical activity $a \propto e^{\mu/kT}$ can be used. In an ideal gas or ideal dilute solution, the chemical potential has a simple logarithmic relationship to concentration and linear in temperature (so that activity is proportional to concentration), and insensitive to the concentrations of other components(154). However, a near-critical mixture is far from ideal – the critical point is precisely where weak cooperative interactions between the many components lead to strong effects (124). Therefore, we expect strong relationships between the chemical potentials of different components, especially when those components modulate Tc.

Many transmembrane proteins have been shown to be modulated by binding to membrane components, prominently including cholesterol(154, 155) and phosphatidylinositol lipids(156), and many other signaling lipids(157). If the chemical potential of some of these components is strongly modulated by concentration changes of other components, binding site occupancy will also vary, and we expect to see changes in protein functions that depend on binding of those components. Recent work by Ayuyan and Cohen has developed sensitive methods for measuring and controlling the chemical potential of cholesterol in the plasma membrane, and found that cholesterol chemical potential varies by ~2kT in different physiologically relevant cellular conditions (154). That amount is certainly adequate to induce substantial changes in binding site occupancy. Work remains to be done to explore if these differences can be attributed in any way to the critical phase transition, but it is exciting to speculate that perturbations that change membrane criticality may act indirectly by impacting the activity of membrane components.

***Criticality coupled to other processes describes a broad array of raft phenomena***: The Ising universal critical phenomena are a good start for understanding membrane heterogeneity, but they are also clearly insufficient to explain all phenomena. As stated above, critical phenomena alone are not expected to give rise to regions of tight clustering or confinement of proteins and lipids, as is sometimes attributed to membrane domains. This said, it is possible that the local membrane environment can impact the conformational states sampled by membrane proteins in ways that facilitate binding through stronger protein binding sites. This type of synergistic effect could underlie a range of cholesterol dependent processes observed at the plasma membrane, including for example the transient pinning observed in studies of membrane protein and lipid dynamics (158).

Another example where there is potential for synergy between criticality and other organizing principles relates to the 'Active Composite Model' proposed by Mayor and coworkers (159). This model posits that the plasma membrane interacts with a heterogeneous cortical actin network, composed of both active and passive components. The passive components largely resemble quenched disorder, as described earlier in this review. The active component is composed of motor driven short actin filaments that can actively drive certain membrane proteins and lipids into close proximity, coupling across membrane leaflets. Considering this model in a critical membrane provides a simple means to correlate domain structure across leaflets without requiring strong interactions such as interdigitation, since the cooperativity inherent in a phase transition can amplify weak couplings that may be present. Moreover, the high susceptibility of a critical membrane allows it to robustly remodel when external forces are applied, including those originating from the actin cortex.

More broadly, we envision that plasma membrane criticality is only one of several organizing principles that contribute to plasma membrane functions. It could be that interactions mediated by curvature and electrostatics superimpose with those mediated via criticality to define the plasma membrane interactome. It is also possible that there is interesting cross-talk between these various interaction modes. For example, studies have shown that the sorting of lipids into curved membranes can be mediated by the binding of curvature-sensing proteins that have preferences for one membrane phase (160). Proteins and peptides can also organize lipids through electrostatics, which in turn can stabilize domains impacted by fluctuations of ordered and disordered phase lipids. The broader implications of this potential cross-talk is largely unexplored, and could give rise to qualitatively new phenomena accessible to cell membranes (161).

## 6. Concluding remarks

While it has long been appreciated that plasma membrane lipids are capable of intriguing, non-ideal behaviors, much of the past literature is clouded by imperfect methods and an incomplete conceptual framework to conceptualize experimental observations. This backdrop led to controversial and often unphysical descriptions of lipid rafts. The past decade or so has brought key advances, including membrane isolations that largely preserve plasma membrane protein and lipid content and super-resolution imaging methods that do not suffer from the same pitfalls that plagued early raft research. Together with this, the membrane community has begun to appreciate the rich phenomena that naturally occur near miscibility critical points, many of which exhibit strong parallels with long-standing observations in both the membrane-biophysics and membrane biology literatures.   Moving forward, the challenge will be to isolate these effects to enable a definitive measurement of the role of criticality in cell membranes, and to explore how these immiscibility-mediated interactions work alongside other physical and biochemical organizing principles to contribute to the rich array of biological functions at the cell surface.

**Figure Captions**

Figure 1: Phase diagram of lipid mixtures. (A) Phase diagrams of three component mixtures are conventionally drawn on an equilateral triangle. The three vertices are pure mixtures of each lipid component, points along the edges are binary mixtures, and points within the triangle contain all three components. Compositions can be read by measuring the perpendicular distance to each edge, which have the property of always summing to 100%. Two examples are shown. (B) Qualitative phase diagram for ternary lipid mixtures of high melting temperature (Tm) lipids, low Tm lipids and cholesterol. Thick red lines indicate boundaries of liquid-solid (So) coexistence and the thick blue line represents the boundary of liquid-disordered (Ld) liquid-ordered (Lo) coexistence. Points along this boundary also indicate the composition of coexisting phases, and the specific compositions in coexistence are indicated by tie-lines (lighter weight lines within binary coexistence regions). The green triangle represents compositions that exhibit all three phases in coexistence, whose compositions are indicated by the three vertices of the triangle. The Ld-Lo coexistence region terminates at a miscibility critical point along the high cholesterol edge, indicated by a yellow star.

Figure 2: Phase diagrams of ternary lipid mixtures exhibit the same overall topology. (A) DiPhytanolPC/DPPC/Chol by fluorescence microscopy at 16°C, redrawn from (41). (B) DOPC/DPPC/Chol by deuterium NMR, from (45). (C) DOPC/DSPC/Chol by fluorescence microscopy and FRET, from (46). (D) DPC/PSM/Chol by FRET, neutron scattering, and DSC, from (47). (E) POPC/PSM/Chol by EPR, from (43). (F) DOPC/eggSM/Chol by AFM, from (48).

Figure 3: Critical systems exhibit universal features that extend beyond the two phase region. (A) Simulation snap-shots of two lattice models. (Top) an Ising model at various reduced temperatures $t$ and compositions $m$ and (middle) a three component model with interactions designed to reproduce ternary GUV phase diagrams, at fixed temperature and various compositions. (Bottom) Compositions of the three-component model are chosen so that the effective reduced temperature $t$ and $m$ of the model match $t$ and $m$ of the corresponding Ising model. Miscibility gaps, tie lines, and the directions of $t$ and $m$ are shown on each schematic phase diagram. (B) Measurements of the divergence of correlation length $\xi$ vs $t$ as the critical point is approached from the one-phase region. (Top) In the two models from panel A. For the three-component model, $t$ is the cholesterol (Chol) content normalized by the Chol content at the critical point. (Middle) In GUVs and GPMVs, varying temperature. In each case, the expected 2D-Ising power law $\xi \propto t^{-1}$ is observed although the proportionality constant differs, resulting in each dataset being plotted with a different y axis scale. (Bottom) Representative GUV and GPMV images near the critical point, data points replotted from (128).

Figure 4: Four Functional Mechanisms Primarily Driven by $L_o/L_d$ Partitioning in a Near-Critical Membrane. Membrane proteins and other components are characterized by how they partition into Lo/Ld domains, as schematically shown. The combination of the partitioning of various

components confers to the systems particular properties that can be used to drive biological function. The four mechanisms described here correspond to those described in greater detail in section 5 of the text. (A) Critical Casimir forces yield an effective attraction between components with like order preference (blue-blue pair), and an effective repulsion between components that prefer opposite lipid order (blue-magenta pair). The strength of these interactions increases rapidly when $t$ is reduced, as $\xi$ becomes large. (B) Clustering a protein that prefers Lo lipids (green) induces a distinct Lo domain due to the high susceptibility of the membrane near the critical point. As a result, other $L_o$-preferring proteins (blue) are recruited to the cluster, and Ld-preferring proteins (magenta) are excluded. Similarly, an Ld domain could be stabilized by clustering an Ld-preferring component. C) The white object is a large protein with two conformations (pentagon/star). One conformation (pentagon) has no order preference and the other (star) prefers an Lo environment. The star conformation becomes more likely as $t$ is decreased because there are large patches of Lo membrane that satisfy its preferred boundary condition. (D) Changes in $t$ also result in changes in the chemical potentials of some membrane components, including components (orange oval) that can bind to a membrane protein (square) as an allosteric modulator, therefore a change in $t$ can induce differences in binding site occupancy and the resulting distribution of protein states.

Figure 1:

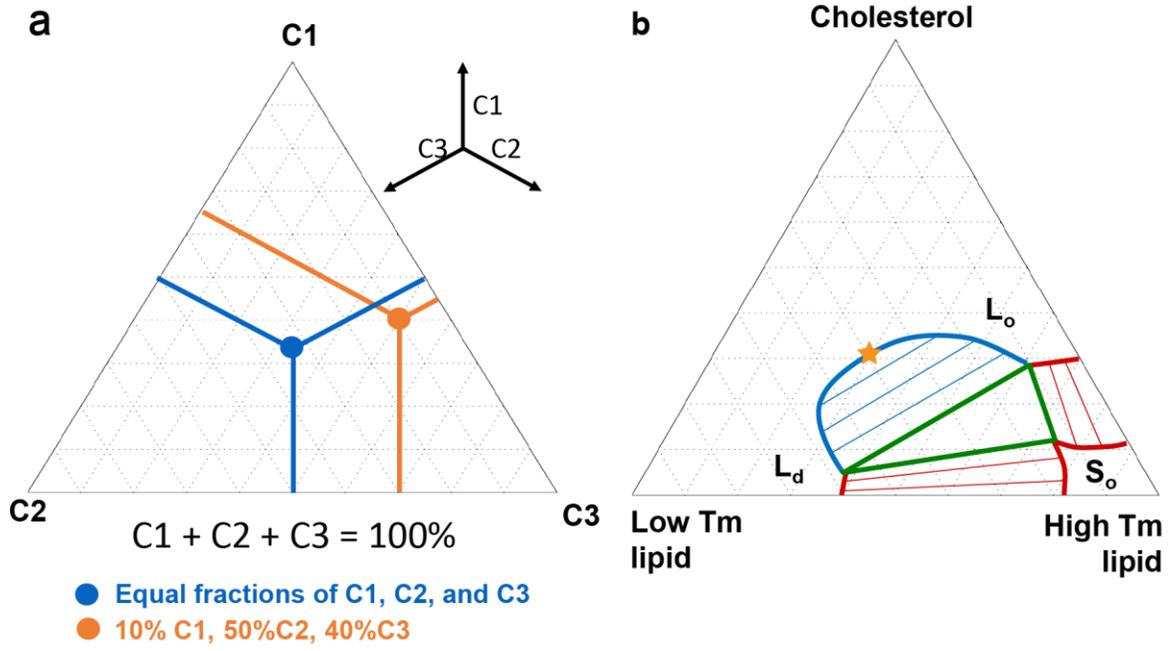

Figure 2 omitted due to pending copyright requests.

Figure 3:

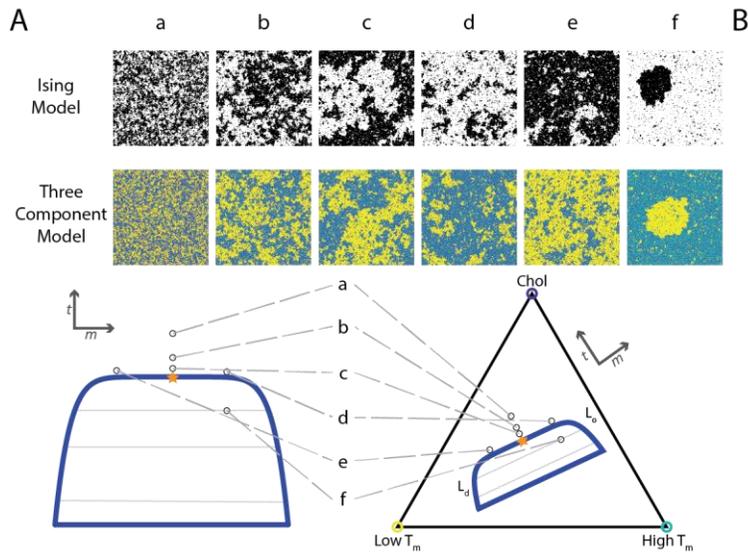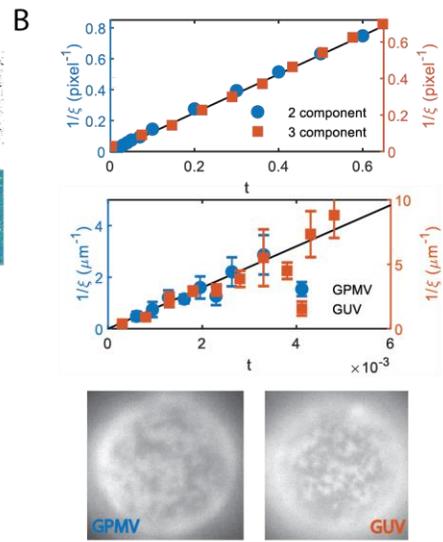

Figure 4:

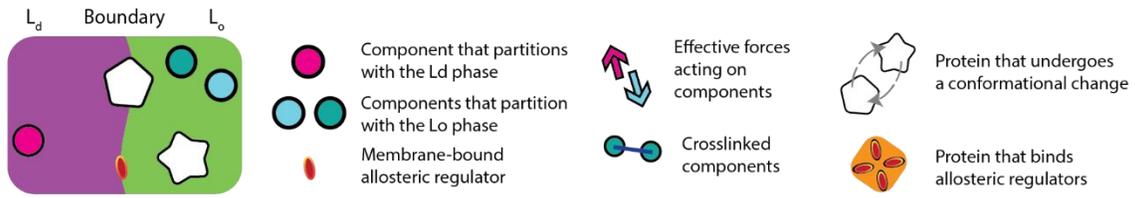

### A. Interactions between proteins

high t/small ξ  
negligible interactions

low t/large ξ  
long-range interactions

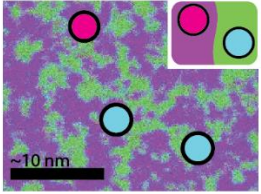 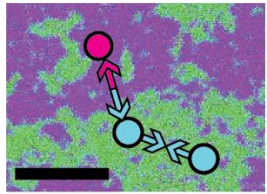

### B. Susceptibility to clustering

high t/small ξ  
minimal sorting

low t/large ξ  
effective sorting

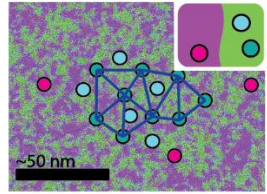 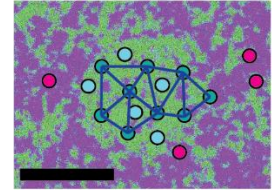

### C. Allosteric regulation of single proteins

high t/small ξ  
state with weak partitioning

low t/large ξ  
state with strong partitioning

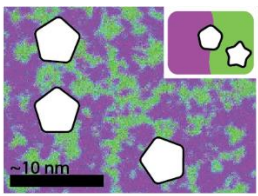 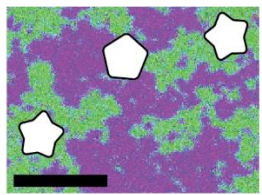

### D. Tuning binding of allosteric regulators

high t/small ξ  
reduced regulator binding

low t/large ξ  
enhanced regulator binding

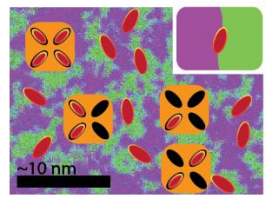 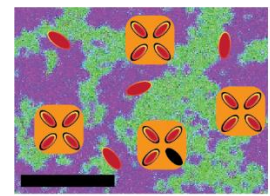